\documentclass[10pt]{article}

\usepackage[margin=1.0in]{geometry}

\usepackage{color}

\usepackage{amsmath}
\usepackage{amsfonts}
\usepackage{amssymb}
\usepackage{caption}           
\usepackage{graphicx}
\usepackage{slashed}            
\usepackage{subcaption}
\usepackage{bbm}                
\usepackage{xspace}				
\usepackage{cite}
\usepackage{float}
\usepackage{dcolumn}
\usepackage{bm}
\usepackage{epstopdf}
\usepackage{feynmp}  
\usepackage[english]{babel}
\usepackage{fancyhdr}
\usepackage{amsmath}
\usepackage{amssymb}
\usepackage{amsfonts}
\usepackage{psfrag}
\usepackage[applemac]{inputenc}
\usepackage{graphicx}

 \newcommand{\vev}[1]{\langle {#1} \rangle}

\def\be{\begin{eqnarray}}
\def\ee{\end{eqnarray}}
\def\bea{\begin{eqnarray*}}
\def\eea{\end{eqnarray*}}
\def\fig#1{Fig.~\ref{#1}}

\begin{document}
\begin{titlepage}
\begin{flushright}
OHSTPY-HEP-T-15-001
\end{flushright}
\vskip.5cm
\begin{center} 
{\huge \bf New Calculations in Dirac Gaugino Models: Operators, Expansions, and Effects \vspace*{0.3cm} } 
\end{center}

\begin{center} 
{\bf  Linda M. Carpenter and Jessica Goodman} 
\end{center}

\begin{center} 

{\it  Department of Physics,  The Ohio State University, Columbus, OH 43210} \\
{\it  lmc@physics.osu.edu, jgoodman@physics.osu.edu } \\

\vspace*{0.1cm}

\end{center}

\vglue 0.3truecm


\begin{abstract}
In this work we calculate important one loop SUSY-breaking parameters in models with Dirac gauginos, which are implied by the existence of heavy messenger fields.  We find that these SUSY-breaking effects are all related by a small number of parameters, thus the general theory is tightly predictive. In order to make the most accurate analyses of one loop effects, we introduce calculations using an expansion in SUSY breaking messenger mass, rather than relying on postulating the forms of effective operators.  We use this expansion to calculate one loop contributions to gaugino masses, non-holomorphic SM adjoint masses, new A-like and B-like terms, and linear terms. We also test the Higgs potential in such models, and calculate one loop contributions to the Higgs mass in certain limits of R-symmetric models, finding a very large contribution in many regions of the $\slashed{\mu}$MSSM, where Higgs fields couple to standard model adjoint fields.

\end{abstract}
\end{titlepage}
\newpage

\section{Introduction}
The first run of the LHC has put new physics studies in an interesting place.  A light Higgs has been discovered, with a mass of 125 GeV \cite{Chatrchyan:2012ufa,Aad:2012tfa}. This raises hope that standard notions of naturalness may still hold, and that new physics may be not be far away.  In this situation supersymmetry (SUSY) is still the leading candidate for beyond standard model (BSM) physics.  However, if SUSY is indeed the correct paradigm, the favored region of SUSY parameter space is different than the expectations going into the first run of the LHC.  Gluino searches in jets plus missing energy channels now place a lower bound of about 1 TeV on gluino and squark masses for MSSM models with standard SUSY mass spectra.  The Higgs mass measurement itself (with heavy stops responsible for the one loop Higgs mass contributions) along with flavor physics measurements demand that the Higgs sector parameters fall within in the decoupling limit.  If the MSSM is indeed the correct picture, we must learn to live with a heavier mass spectrum and a larger dose of fine tuning.

However, interesting extensions to the MSSM scheme exist, among these are  R-symmetric models.  These models have Dirac gauginos, which require the existence of new chiral superfields which are adjoints under the standard model (SM) gauge groups.  The adjoints are the fields that `marry' the gluinos with a Dirac mass \cite{Fayet:1978qc,Hall:1990hq}.  These models present many interesting phenomenological features, and are known to be relatively unconstrained by LHC's first run \cite{Kribs:2012gx}.  Most models of this type have non-unified gaugino masses, which allow for a more complex sparticle spectra.  They naturally predict heavy gluinos, suppressing gluino pair production, squark pair production and squark gluino associated production.  In addition, they allow for complex multi-particle decay chains \cite{Fox:2014moa}.

In a very predictive implementation known as Supersoft, R-symmetric gaugino masses are generated at the 10 TeV scale.  Scalar sparticle masses are then generated one loop level down and are finite, cut off by diagrams containing scalars adjoint fields \cite{Fox:2002bu}.  Most implementations of R-symmetric models are built in gauge mediated scenarios \cite{Dine:1993yw,Dine:1994vc}, where SUSY breaking is communicated to SM sparticles through loop interactions with messenger fields. The mass spectra of R-symmetric models are highly dependent on various one loop effects.  For example, SUSY breaking masses for scalar adjoints are generated at the same order as gluino masses, and through similar mechanisms.  These scalar masses drastically effect the viability of models, adjoint masses squared must themselves be large and positive to be phenomenologically viable, and they feed into the squark, Higgs, and slepton mass spectra.  Thus we see that understanding all one-loop SUSY breaking operators induced by the messenger sector is extremely important.

One place where fully comprehending the effects of the messenger sector is crucial is the Higgs-sector. Analyses of the Higgs sector of R-symmetric models are intricate, and heavily rely on one-loop effects.  One challenge for R-symmetric models is to achieve a Higgs mass of 125 GeV while maintaining a viable sparticle spectrum. It is well known that in the MSSM, the tree-level Higgs mass may not exceed the mass of the Z.  Large one loop corrections must follow from the stops, the only fields with appreciable coupling to the Higgs.  These large corrections are aided by  large mixing in the stop sector due to large A-terms. R-symmetric models, however, lack such terms, and therefore new mechanisms must be invoked to raise he tree level Higgs mass.  One such mechanism is the so called $\mu$-less MSSM or $\slashed{\mu}$MSSM.  These models include superpotential terms, allowed by all symmetries, that couple the Higgs fields to the new SM adjoints \cite{Nelson:2002ca}. These terms raise the tree level Higgs mass through the introduction of new quartic couplings. We will show that SUSY breaking effects in these models will produce new, large one loop Higgs-mass-corrections similar to stop loops which drastically effect Higgs phenomenology.  In addition, as we demonstrate, several R-breaking, loop level operators will be generated once electroweak symmetry is broken including A-like and b-terms and linear terms. The dynamics of the Higgs, and thus the viability of many R-symmetric models, will depend heavily on these SUSY breaking effects.

In this work we focus on one loop SUSY breaking effects which make important contributions to low energy spectra in R-symmetric models.  Past calculations of scalar adjoint masses have been complicated by their attempted identification with effective operators in the superpotential. For example, it was first assumed that either the real or imaginary component of scalar adjoint fields must be tachyonic. Upon further calculation in specific models, this was found not to be the case, however, confusion remained as to the exact operator responsible for the mass.  We find that individual operators do not capture the physics to all orders in SUSY breaking.  Instead of relying on effective operators we use the technique of calculating one loop effects by using a simple expansion of  messenger propagators in powers of the SUSY breaking parameters.  We can then easily capture SUSY breaking effects to all orders.

We present a calculation to any order of the one loop SUSY breaking masses of scalar adjoints in R-symmetric models.  We also use this technique to calculate the size of SUSY-breaking operators in the Higgs sector which will be necessary for successful electroweak symmetry breaking and Higgs mass prediction.  We will introduce new A-like and b-like terms and analyze the Higgs potential in favored regions of parameter space.

In Section 2, we review the one loop calculation of Dirac masses in R-symmetric models. We present the history of one loop mass calculation for SM adjoint scalars, and present a new calculation using a power expansion in terms of SUSY breaking parameters.  In section 3 we present one loop calculations for various SUSY-breaking operators which evolve in Dirac gaugino models. This includes calculations of trilinear A-like terms involving the adjoint scalar, b-terms for Higgses and linear terms. In section 4 we review the full Higgs-sector scalar potential with general allowed operators in R-symmetric models and explore minima characteristic of supersoft models.  In section 5 we present one loop contributions to the Higgs masses for a sub-set of models in the $\slashed{\mu}$MSSM and demonstrate that these contributions may lead to large corrections to the Higgs mass. Section 6 concludes.

\section{Dirac Gauginos, Supersoft and a History of Loop Masses}

Dirac gaugino masses may be generated from a complete SUSY-breaking model through higher dimensional operators.  They require a set of new chiral superfields which are adjoints under the standard model gauge groups.  Many implementations of R-symmetric models rely on dimension 4 superpotential operators which involve effective interactions between the gaugino, new adjoint, and a hidden sector $U(1)$ field which gets a D term vev.  The relevant operator, known as the supersoft operator \cite{Fox:2002bu} is
\begin{equation}
 W_{\text{ssoft}}=\int d^2 \theta \zeta_j\frac{W'_{\alpha} W^\alpha_j A_j}{\Lambda} \sim \int d^2 \theta \theta^2\frac{\zeta_j D_\alpha}{\Lambda}\lambda^\alpha_j \Psi_{A_j},
\end{equation}
where W$'$ is the $U(1)'$ field strength,  W is the standard model field strength (either $SU(3), SU(2), \text{ or } U(1)$) and A a standard model adjoint.  Once the $U(1)'$ field is set to its SUSY breaking vev, the operator becomes a Dirac mass term mixing the gaugino with the adjoint fermion. Here the coupling $\zeta$ may be different for each SM gauge group (denoted by index j), therefore R-symmetric models in general predict that gaugino masses are non-unified.

Supersoft SUSY breaking, so named because it introduces no new divergences to the soft SUSY breaking parameters, extends the
low energy gauge sector of the MSSM by introducing an adjoint superfield for each gauge symmetry.
Dirac gaugino masses are then generated through the supersoft operator once some hidden $U(1)'$ obtains a D-term.

In gauge mediated UV completions of supersoft models, the gaugino mass is generated from interactions with a messenger sector which couples both to the dynamical SUSY breaking sector, and to the standard model fields. These are general gauge mediated completions where gaugino and MSSM scalar masses vanish as SM gauge couplings are set to zero \cite{Meade:2008wd}\cite{Benakli:2008pg}   To facilitate the gaugino mass, the messengers must be charged both under the hidden sector U(1), and the standard model gauge group. This class of models thus belongs under the taxonomy of `semi-direct' gauge mediated models, the messengers are charged under the hidden sector gauge groups, but do not participate in the SUSY breaking itself \cite{Seiberg:2008qj}.  We can see how $W_\text{ssoft}$ is generated, for example, by considering a toy superpotential containing
the messengers $\phi$ and $\bar{\phi}$, which are charged under the $U(1)^{'}$ and are also fundamentals or anti-fundamentals under the SM gauge groups.  This superpotential contains a supersymmetric mass for the messengers, as well as a Yukawa-like interaction between the messengers and the chiral adjoint,
\begin{equation}
W \sim \lambda_i \phi A_i \bar{\phi} +m_{\phi} \phi \bar{\phi}.
\label{eq:gmsbW}
\end{equation}
As the messengers are charged under the broken $U(1)$, they acquire a diagonal non-supersymmetric mass contribution proportional to the $U(1)^{'}$ D-term.  Given these interactions, the gaugino mass follows at one loop level,
\begin{figure}[H]
          \centering
	    \begin{fmffile}{Diracmass}
	        \begin{fmfgraph*}(120,50)
	       	     \fmfstraight
	        		\fmfleft{i1,i2}
	      		 \fmfright{o1,o2}
             		  \fmf{fermion}{i1,v1}
                 	  \fmf{fermion,label=$\phi$}{v2,v1}
                		 \fmf{fermion,label=$\bar{\phi}$,label.side=left}{v2,v3}
                 	\fmf{fermion}{o1,v3}
                 	\fmf{phantom}{i2,v4,o2}
               \fmffreeze
                \fmf{photon}{i1,v1}
                \fmf{scalar, label=$\phi^\dagger$,left=0.55,tension=0.6}{v1,v4}	
                 \fmf{scalar,label=$\phi$, left=0.55,tension=0.6}{v4,v3}	
                  \fmflabel{$\lambda$}{i1}
	        \fmflabel{$A$}{o1}
		\fmfv{decoration.shape=cross,decoration.angle=0,decoration.size=5thick}{v2}
		\fmfv{decoration.shape=cross,decoration.angle=0,decoration.size=5thick}{v4}
		\fmflabel{$D$}{v4}
		\fmflabel{$m_\phi$}{v2}
	        \end{fmfgraph*}
	    \end{fmffile} \vspace{.25cm}
	    \caption{1-loop diagram leading to Dirac gaugino masses.}
\label{fig:DiracMass}
\end{figure}
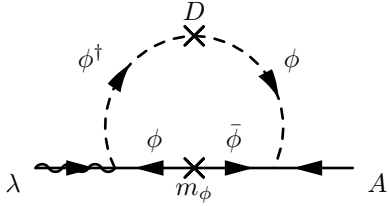
\noindent giving a Dirac mass,
\be
m_D \sim \frac{g_{\text{SM}} g'\lambda_{\phi}}{16\pi^2}\frac{D}{m_\phi}
\ee
as we expect in accordance with the operator in eqn 1. An additional consequence of $W_\text{ssoft}$ is the shift in the standard model D-terms,
\begin{eqnarray}
D\rightarrow m_D(A+A^{\dagger})+\Sigma_i gQ_i^*TQ_i
\end{eqnarray}
which alters the D-term generated Higgs quartic as well as generates a mass for the real part of the adjoint field.
In fact, for a heavy adjoint and in the absence of a supersymmetric Majorana
mass term for the adjoint, one finds that the D-term Higgs quartic vanishes.  However, the models we consider below are variants
of the $\mu$-less supersymmetric standard model \cite{Nelson:2002ca} and therefore contain trilinear superpotential operators
coupling the Higgs superfields to the adjoint superfields.  These superpotential terms generate tree level contributions to the Higgs
quartic term as well as additional loop contributions to the Higgs mass similar to those gained from stop/top loops as we will see below.

We will see that in addition to the Dirac mass, the SUSY breaking contribution of the messengers also induce a variety of other one loop SUSY-break effects into the SM sector-from scalar masses to A and b-like terms.  Some of these will be of the same order as the gaugino mass, while some will be suppressed.  It is important to note that once the parameters in SUSY breaking and messenger sectors are chosen, all of the one loop SUSY breaking contributions are then fixed.  We will first consider the effect of the messenger on the chiral adjoint fields.

\subsection{High Energy Models}
In order to shed light on operators which result from effects of messengers, we will now give a simple example of a SUSY breaking sector where messengers develop a D-term mass. One of the simplest hidden sectors which results in a broken  $U(1)^{'}$ D-term is a modified O'Raifeartaigh model \cite{Dine:2007dz}.
Consider the following hidden sector superpotential with a gauged U(1) symmetry

\be
W= \lambda X (\psi_{+}\psi_{-}- \mu^2) + m_1\psi_{+}Z_{-} + m_2 \psi_{-}Z_{+}+ W^{'}W^{'}
\ee
where the subscripts indicate $U(1)^{'}$ charges.  We see that the fields $\psi$ get $U(1)^{'}$
breaking vevs

\be
\psi_{+}^2=\frac{m_2}{m_1}\phi_{-}^2
\ee
\be
\psi_{-}= \sqrt{\frac{m_1}{m_2}\mu^2-m_1^2} \nonumber
\ee
While the field X as well as the Z's get SUSY-breaking F terms of order $m/\lambda$.

The D term is nonzero as long and $m_1$ is unequal to $m_2$, and is given by
\be
D = g^{'}(\frac{m_1}{m_2}\mu^2-m_1^2)(\frac{m_2}{m_1}-1)
\ee

The SUSY breaking is communicated  through addition of a set of messengers in the fundamental representation of the SM
gauge groups, and which are also charged under the $U(1)^{'}$.
The simplest messenger sector we may write is again

\be
W_\phi= m_\Phi \Phi\overline{\Phi}+ y_i\overline{\Phi}A\Phi
\ee
This messenger sector preserves a messenger parity, that is, the superpotential is symmetric under the interchange $\phi \rightarrow \overline{\phi}$.  A more general messenger sector may contain multiple sets of messengers $\Phi_i$ with their own couplings and Dirac masses,
\be
W_\phi= m_{ij} \Phi_i\overline{\Phi_j}+ y_{ij}\overline{\Phi_i}A\Phi_j
\ee
Here, depending on masses an couplings, a messenger parity $\Phi_i \rightarrow \overline{\Phi_j}$ is not necessarily present.  To cancel anomalies each  fundamental must come with an anti-fundamental of opposite charge,  however, each \textbf{set} of fundamental/antifundamental messengers may have different magnitude of charges under the hidden sector U(1).  In the simplest completions $m_{ij}$'s are given the by vevs of a set of dynamical fields.

\subsection{Adjoint Masses}
One persistent question in the study of Dirac gauginos has been about the operators involved in generating the adjoint mass terms.  These masses may have drastic effects on supersymmetric spectra.  In supersoft models, it is the real part of the adjoint field that couples to MSSM scalars, and appears in 2-loop  gauge mediated diagrams which determines their masses.  In addition, parts of the adjoint multiplet may themselves be observable in colliders, and have effects on particle decay chains.  It is therefore  crucial for a  predictive theory to correctly capture the adjoint mass contributions.

 $W_\text{ssoft}$ will contribute to the mass of the real part of the scalar in the adjoint superfield,
while the imaginary piece will remain massless assuming no other explicit contribution to the adjoint mass.
However, it was noted in the original formulation of supersoft models, \cite{Fox:2002bu}, that once $W_\text{ssoft}$ is allowed, there is no symmetry forbidding the `lemon twist' operator,
\begin{equation}
W_\text{LT} \sim \zeta'\frac{W'W'AA}{\Lambda^2},
\end{equation}
which leads to a holomorphic adjoint
scalar mass on the order of $m_D^2$ once the D-terms are inserted.  This operator will in fact lead opposite sign masses for the real and imaginary parts of the adjoint, and thus to a large tachyonic mass for the imaginary
component of the adjoint scalar field.

It was suggested in \cite{Carpenter:2010as}, that a remedy for this problem comes from considering the full set of one
loop diagrams which contribute to the scalar adjoint masses.  While the operator above contributes to the holomorphic mass of A, in fact, due to messenger loops, there are several diagramatic contributions both to the
holomorphic mass and the non-holomorphic mass of A.  It is not difficult to show that loops linear in D (or with any odd power of D insertions) will cancel
since the messengers have opposite charges under the SM.  Thus, one must consider diagrams of order $D^2$ as shown below.  The first diagram in
Fig.~\ref{fig:quadraticD} contributes to the mass of $AA^\dag$ while the second and third diagrams contribute to both the holomorphic and non-holomorphic mass for A.  \\
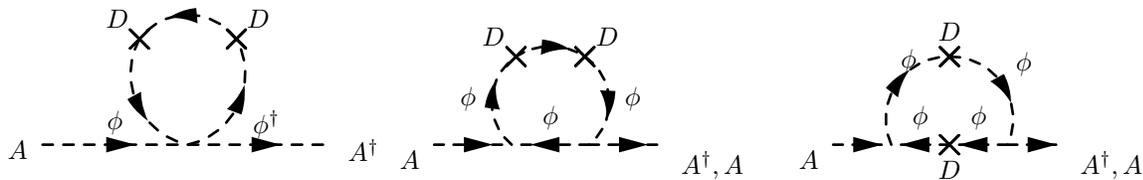
\begin{figure}[H]

\begin{subfigure}[b]{.3\textwidth}
\centering
\begin{fmffile}{quadraticDquartic}
	        \begin{fmfgraph*}(110,40)  
	     \fmfstraight
	        \fmfleft{i1,i2}
	        \fmfright{o1,o2}
                 \fmf{scalar}{i1,v1,o1}
                 \fmf{phantom}{i2,v2,v3,o2}
               \fmffreeze
                \fmf{scalar, label=$\phi$,right=0.5,tension=0.3}{v2,v1}	
               \fmf{scalar,label=$\phi^\dagger$, right=0.5,tension=0.3}{v1,v3}	
                \fmf{scalar, right=0.5,tension=0.6}{v3,v2}	
                  \fmflabel{$A$}{i1}
	        \fmflabel{$A^\dagger$}{o1}
\fmfv{decoration.shape=cross,decoration.angle=0,decoration.size=5thick}{v2}
\fmfv{decoration.shape=cross,decoration.angle=0,decoration.size=5thick}{v3}
\fmflabel{$D$}{v2}
\fmflabel{$D$}{v3}
	        \end{fmfgraph*}
	    \end{fmffile}
	    \end{subfigure}
~
     \begin{subfigure}[b]{.3\textwidth}
	     \begin{fmffile}{quadDcubic}
        \begin{fmfgraph*}(110,40)
                   \fmfleft{i1,i2}
                \fmfright{o1,o2}
                   \fmflabel{$A$}{i1}
                  \fmflabel{$A^\dag , A$}{o1}
                  \fmfstraight
                   \fmf{scalar}{i1,v1}
                   \fmf{scalar,label=$\phi$}{v4,v1}
                    \fmf{scalar}{v4,o1}
                    \fmffreeze
                   \fmf{phantom,tension=.5}{i2,v2,v3,o2}
                   \fmf{scalar,label=$\phi$,left=0.5,tension=.1}{v3,v4}
                   \fmf{scalar,left=.3,tension=.1}{v2,v3}
                  \fmf{scalar,label=$\phi$,left=0.5,tension=.1}{v1,v2}
                  \fmffreeze
                  \fmfv{decoration.shape=cross,decoration.angle=0,decoration.size=5thick}{v2}
		\fmfv{decoration.shape=cross,decoration.angle=0,decoration.size=5thick}{v3}
                   \fmflabel{$D$}{v2}
                   \fmflabel{$D$}{v3}
	        \end{fmfgraph*}
	   \end{fmffile}
    \end{subfigure}
~
     \begin{subfigure}[b]{.3\textwidth}
	     \begin{fmffile}{quadDcubic1}
        \begin{fmfgraph*}(110,40)
                \fmfleft{i1,i2}
                	\fmfright{o1,o2}
                 \fmflabel{$A$}{i1}
                 \fmflabel{$A^\dag,A$}{o1}
                 \fmfstraight
                 \fmf{scalar}{i1,v1}
                 \fmf{scalar,label=$\phi$}{v3,v4,v1}
                 \fmf{scalar}{v3,o1}
                 \fmffreeze
                 \fmf{phantom,tension=.5}{i2,v2,o2}
                 \fmf{scalar,label=$\phi$,left=0.55,tension=0.1}{v2,v3}
                 \fmf{scalar,label=$\phi$,left=0.55,tension=0.1}{v1,v2}
                 \fmffreeze
                 \fmfv{decoration.shape=cross,decoration.angle=0,decoration.size=5thick}{v2}
		\fmfv{decoration.shape=cross,decoration.angle=0,decoration.size=5thick}{v4}
                 \fmflabel{$D$}{v2}
                 \fmflabel{$D$}{v4}
	   \end{fmfgraph*}
	   \end{fmffile}
    \end{subfigure}
\\
\caption{\label{fig:quadraticD} Contributions to the adjoint masses from messenger loops.  There are additional contributions with $\phi$
replaced by $\bar{\phi}$.}
\end{figure}

However, matching the non-holomorphic mass contributions to an operator has proven harder. It was first suggested \cite{Carpenter:2010as} that the non-holomorphic masses were contained in the operator
\be
K= \int d^4 \theta \frac{W^{'}DV^{'}AA^{\dagger}}{\Lambda^2}+ h.c.
\ee
However, this operator is not supergauge invariant and cannot capture physics of the diagrams above.  In fact, in models
with messenger parity the $\mathcal{O}(D^2)$ diagrams in Fig.~\ref{fig:quadraticD} will vanish though the $\mathcal{O}(D^4)$
will be non-zero.  It was later suggested \cite{Csaki:2013fla}
that the correct operator owing to the non-holomorphic masses is
\be
K= \int d^4 \theta \frac{1}{\Lambda^2}(\psi^{\dagger}  e^{qV}\psi + \overline{\psi}^{\dagger}e^{-qV} \overline{\psi})\text{Tr} A^{\dagger}A+ h.c.
\label{eq:Csaki}
\ee
where $\psi$ and $\overline{\psi}$ are the fields responsible for breaking the hidden $U(1)'$ and thus generating the non-zero D-term.
Furthermore, it was argued in \cite{Csaki:2013fla} that this operator is always generated at two loops while in the presence of messenger parity
violation it is generated at one loop.  There are some issues with this formulation as well.  While Eq~\ref{eq:Csaki} applies to a non-holomorphic adjoint mass to order $D^2$, one can
imagine generating the adjoint mass through one loop diagrams with multiple D-term insertions along the messenger propagators.  This suggests summing the series expansion
of the messenger propagators to obtain corrections to the adjoint mass to all orders in D.  We will discuss this in more detail in the next section.

\subsection{Messenger Propagators and SUSY-breaking Power Expansion}
All SUSY-breaking effects are ultimately fed to MSSM fields through messenger loops. In order to fully calculate these
effects we must understand the effect of SUSY breaking on the messenger propagators.  Here we propose a formalism to
calculate all loop level SUSY-breaking mass parameters  to any order in the SUSY breaking parameter D.

Consider a toy model containing a single set of messengers $\phi$ and $\bar{\phi}$ which are a fundamental and anti-fundamental under SM gauge groups, and have opposite charges under the hidden sector U(1). The messengers have a supersymmetric mass and coupling to the adjoint field as in Eq. \ref{eq:gmsbW}.
Since the messenger fields are charged under the hidden $U(1)'$ with charges $\pm q$, a non-holomorphic messenger mass in generated in the a scalar potential through D-terms

\begin{equation}
V \sim g'q(|\phi|^2-|\bar{\phi}|^2)D
\end{equation}
These terms generate corrections to the messenger two-point function. We may write the messenger propagator
 as a series expansion in terms of insertions of the SUSY breaking D-term as seen in Fig.~\ref{fig:propseries}.
\begin{figure}[H]
\begin{equation}
\begin{subfigure}[H]{.25\textwidth}
	\centering
		\begin{fmffile}{fullprop}
	        		\begin{fmfgraph*}(100,30)  
	     			\fmfstraight
	        			\fmfleft{i1}
	        			\fmfright{o1}
                 		\fmf{dashes}{i1,v1,o1}
				\fmfblob{.15w}{v1}
	        		\end{fmfgraph*}
	    	\end{fmffile}
\end{subfigure}	
=
\begin{subfigure}[H]{.25\textwidth}
	\centering
		\begin{fmffile}{fullprop1}
	        		\begin{fmfgraph*}(100,30)  
	     			\fmfstraight
	        			\fmfleft{i1}
	        			\fmfright{o1}
                 		\fmf{dashes}{i1,v1,o1}
		                 \fmfv{label=$g'qD$,label.angle=90,decoration.shape=cross,decoration.angle=0,decoration.size=5thick}{v1}
	        		\end{fmfgraph*}
	    	\end{fmffile}
\end{subfigure}	
+
\begin{subfigure}[H]{.25\textwidth}
	\centering
		\begin{fmffile}{fullprop2}
	        		\begin{fmfgraph*}(100,30)  
	     			\fmfstraight
	        			\fmfleft{i1}
	        			\fmfright{o1}
                 		\fmf{dashes}{i1,v1}
		                 \fmfv{label=$g'qD$,label.angle=90,decoration.shape=cross,decoration.angle=0,decoration.size=5thick}{v1}
		                 \fmflabel{$g'qD$}{v1}
		                 \fmf{dashes}{v1,v2}
		                 \fmfv{label=$g'qD$,label.angle=90,decoration.shape=cross,decoration.angle=0,decoration.size=5thick}{v2}
		                 \fmflabel{$g'qD$}{v2}
		                 \fmf{dashes}{v2,o1}
	        		\end{fmfgraph*}
	    	\end{fmffile}
\end{subfigure}	
+ ...	\nonumber 	
\end{equation}
\caption{\label{fig:propseries} Corrections to messenger propagator due to non-zero D term.}
\end{figure}
\noindent For any messenger with supersymmetric mass m and SUSY breaking mass D we may re-sum the series to get the full messenger propagator,
\begin{eqnarray}
\frac{1}{p^2-m^2}+\frac{g'qD}{(p^2-m^2)^2}+\frac{g'qD^2}{(p^2-m^2)^3}+ ... =\frac{1}{p^2-m^2-g'qD}.
\end{eqnarray}

We see that, as we expect, the full effect of the series summation of D-terms is to shift the diagonal mass-squared of the messengers by $+ / -$ D for the fundamental/anti-fundamental.   It is this non-holomorphic mass of the messengers that feeds down, generating all SUSY breaking mass parameters.  We see that to calculate one loop-effects at any order in the SUSY-breaking parameter D, we may simply draw a one loop diagram with messengers and expand the propagator to the desired order.  This procedure is quite general, it bypasses the need for guessing operators, and it provides a more unified framework for calculating SUSY-breaking parameters at different orders in SUSY-breaking.

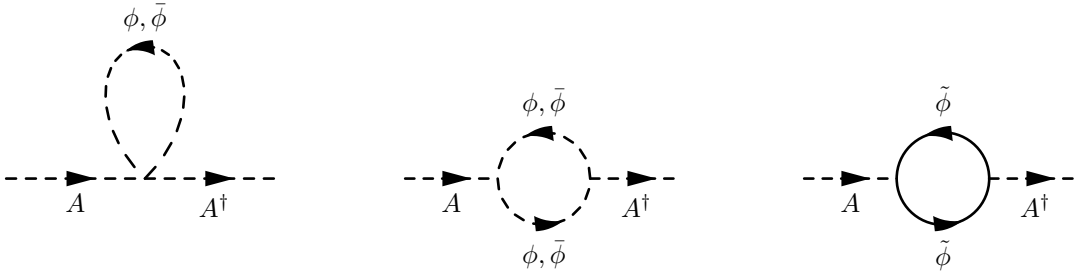
\begin{figure}[H]
         \begin{subfigure}[b]{.3\textwidth}
         \centering
	    \begin{fmffile}{fullprop4point}
	        \begin{fmfgraph*}(105,105)
                   \fmfleft{i1}
                   \fmfright{o1}
                   \fmf{scalar,label=$A$}{i1,v1}
                   \fmf{scalar,label=$\phi ,, \bar{\phi}$,right,tension=.7}{v1,v1}
                   \fmf{scalar,label=$A^\dag$}{v1,o1}
	        \end{fmfgraph*}
	    \end{fmffile}
           \end{subfigure}
~
         \begin{subfigure}[b]{.3\textwidth}
         \centering
	    \begin{fmffile}{fullprop3point}
	        \begin{fmfgraph*}(105,105)
                   \fmfleft{i1}
                   \fmfright{o1}
                   \fmf{scalar,label=$A$}{i1,v1}
                   \fmf{scalar,label=$\phi ,,\bar{\phi}$,right,tension=.5}{v2,v1,v2}
                   \fmf{scalar,label=$A^\dag$}{v2,o1}
	        \end{fmfgraph*}
	    \end{fmffile}
           \end{subfigure}
~
        \begin{subfigure}[b]{.3\textwidth}
        \centering
	    \begin{fmffile}{fullprop3pointF}
	        \begin{fmfgraph*}(105,105)
                   \fmfleft{i1}
                   \fmfright{o1}
                   \fmf{scalar,label=$A$}{i1,v1}
                   \fmf{fermion,label=$\tilde{\phi}$,right,tension=.5}{v2,v1,v2}
                   \fmf{scalar,label=$A^\dag$}{v2,o1}
	        \end{fmfgraph*}
	    \end{fmffile}
	  \end{subfigure}
\caption{\label{fig:resumAmass} Diagrams contributing to the soft masses for the adjoint field due to messenger loops.}
\end{figure}

As an example of this procedure, we now present a calculation of SUSY-breaking adjoint scalar masses.
One can then express the exact one loop contributions to the adjoint scalars, A, as the sum of diagrams in Fig. \ref{fig:resumAmass}.
Re-summing the propagators, we note that  we must now include the contribution from the fermion messengers
in order to properly cancel the $\mathcal{O}(D^0)$ terms. \vspace{.5cm}

Summing the contributions from all three diagrams we find the full D dependent correction to the non-holomorphic adjoint mass to be
\begin{eqnarray}
-i\delta m_A^2 &=& \frac{iy^2}{16\pi^2}\Biggl[(2m^2+D)\left[\text{ln}\left(\frac{\Lambda^2+m^2+D}{\Lambda^2+m^2}\right)-\text{ln}\left(\frac{m^2+D}{m^2}\right)\right] \nonumber \\
&+&(2m^2-D)\left[\text{ln}\left(\frac{\Lambda^2+m^2-D}{\Lambda^2+m^2}\right)-\text{ln}\left(\frac{m^2-D}{m^2}\right)\right] \nonumber \\
&+&\frac{m^2(m^2+D)}{\Lambda^2+m^2+D} +\frac{m^2(m^2-D)}{\Lambda^2+m^2-D}-\frac{2m^4}{\Lambda^2+m^2}\Biggr]
\end{eqnarray}
It is clear that in the supersymmetric limit $D \rightarrow 0 $, this correction vanishes as expected.  We may now expand in powers of the parameter $D/m^2$ to the desired order.   Expanding, we find that the first non-zero contribution to the
non-holomorphic adjoint mass is
\be
\delta m_A^2 \sim \frac{y^2}{240\pi^2}\left(\frac{5}{m^2}\left(\frac{D}{m}\right)^4+\frac{4}{m^4}\left(\frac{D}{m}\right)^6\right)+\frac{1}{m^6}\mathcal{O}\left(\frac{D}{m}\right)^8
\ee
The leading contribution is $\mathcal{O}(D^4)$. This agrees with the results of  \cite{Csaki:2013fla}.  In this simple model with a messenger parity, we expect an accidental cancelation of the SUSY-breaking mass at order $D^2$ leaving $D^4$ the first non-zero order.  This procedure can capture the effects of multiple sets of messengers fairly simply.  In that case one simply calculates the same diagrams for each set of messengers. Each messenger may have a different ratio $D/m^2$ depending on its U(1) charge and supersymmetric mass. If the messenger sector does not contain a messenger parity, one finds SUSY breaking adjoint masses at order $D^2$.

In principle, for non-holomorphic adjoint masses, we are calculating the wave-function renormalization
\be
K \sim \int d^4 \theta Z_A A^{\dagger} A
\ee
expanding in powers of SUSY breaking parameter. This procedure leads to a calculation of the messenger mass contribution at any order in SUSY breaking field D. We note that one cannot capture
these effects by attempting to use the one-loop wave-function renormalization techniques\emph{ ala} Giudice and Rattazzi  which can only yield the initial term the expansion \cite{Giudice:1997ni}.  The SUSY breaking contributions do not correspond any one operator yet proposed to explain the one loop results.

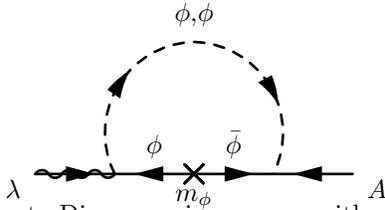
\begin{figure}[H]
        	  \centering
	 	   \begin{fmffile}{DiracmassFullProp}
	        		\begin{fmfgraph*}(120,50)
	       	     		\fmfstraight
	        			\fmfleft{i1,i2}
	      		 	\fmfright{o1,o2}
             		  	\fmf{fermion}{i1,v1}
                 	  	\fmf{fermion,label=$\phi$}{v2,v1}
                		 	\fmf{fermion,label=$\bar{\phi}$,label.side=left}{v2,v3}
                 		\fmf{fermion}{o1,v3}
                 		\fmf{phantom}{i2,v4,o2}
              			 \fmffreeze
               			 \fmf{photon}{i1,v1}
               			 \fmf{scalar,left=0.55,tension=1}{v1,v4,v3}	
                			\fmflabel{$\phi$,$\bar{\phi}$}{v4}
                  		\fmflabel{$\lambda$}{i1}
	        			\fmflabel{$A$}{o1}
				\fmfv{decoration.shape=cross,decoration.angle=0,decoration.size=5thick}{v2}
				\fmflabel{$m_\phi$}{v2}
	        			\end{fmfgraph*}
	    			\end{fmffile}\\
	    \caption{1-loop diagram leading to Dirac gaugino masses with corrected messenger propagator.}
\label{fig:DiracMassFullProp}
\end{figure}

We note that any one loop result in the supersoft formalism, including the gaugino mass itself, may be obtained through the procedure
of expanding the messenger propagator.  Recall the gaugino mass receives a one loop contribution from each messenger, Fig~ \ref{fig:DiracMassFullProp}. \vspace{.25cm}
\noindent Summing up the contributions from both $\phi$ and $\bar{\phi}$ using the fully corrected messenger propagator, one finds
\begin{eqnarray}
-i\delta m_\lambda &=& \frac{-iygm\sqrt{2}}{32\pi^2}\frac{1}{g'D}\Biggl[(m^2+g'D)\left(\text{ln}\left[\frac{\Lambda^2+m^2+g'D}{\Lambda^2+m^2}\right]-\text{ln}\left[\frac{m^2+g'D}{m^2}\right]\right)
\nonumber \\ &+&
(m^2-g'D)\left(\text{ln}\left[\frac{\Lambda^2+m^2-g'D}{\Lambda^2+m^2}\right]-\text{ln}\left[\frac{m^2-g'D}{m^2}\right]\right)\Biggl]\nonumber \\
&=&\frac{iyg\sqrt{2}}{32\pi^2}\left(\frac{g'D}{m}+\frac{1}{6m^2}\left(\frac{g'D}{m}\right)^3\right)+\frac{1}{m^4}\mathcal{O}\left(\frac{g'D}{m}\right)^5
\end{eqnarray}
Recall that for a set of messengers $\phi$ and $\bar\phi$, there are two diagrams that contribute to this process: one with the un-barred messenger as the scalar propagator and one with the barred messenger as the scalar propagator.  We find the first term in the expansion is proportional to D noting that at this order the two diagrams have the same sign.  That is one picks up a negative sign at both the D-term insertion and at the gauge
vertex.  It should also be noted that all terms proportional to $D^{2n}$ vanish as the contributions from the $\phi$ diagrams cancel those from
$\bar{\phi}$ diagrams.

These correction may be expressed in operator form,
\be
K=\Sigma_n \int d^4\theta \frac{(D^2{\overline{D}}^2 V^{'})^{2n}}{\Lambda^{2n}} \frac{W^{'}DV A}{\Lambda}
\ee
which is still R-preserving.  This suggests that the full operator correspondence to one loop effects involves an exponentiation by summing a series of operators.  We note that as this work was in completion,the work \cite{Nelson:2015cea} appeared, which contains similar operators for new Supersoft terms.

\section{Generating A-like, b-like, and Linear Terms}

We note that, in addition to SUSY-breaking adjoint masses,  messenger effects result in many operators at one loop level, some of which pertain directly to the Higgs sector. We will first consider A-term like SUSY-breaking parameters. Indeed, in Dirac-gaugino models models, one may imagine writing an A-like operator in the superpotential
\be
W= \int d^2 \theta {\delta}_s \frac{W'W'}{\Lambda^3}S H_u H_d + \int d^2 \theta {\delta}_{A} \frac{W'W'}{\Lambda^3}{A} H_u H_d
\ee
which is an A-like term of order ${D^2}/{\Lambda^3}$.  This term is a coupling between the weak-sector adjoints of a Dirac-gaugino model and the Higgses, here S is the singlet while A is the SU(2) adjoint.

In the absence of any further structure, such an operator may result from one loop diagrams resulting from Higgs and adjoint couplings to messengers.  While the adjoint field couples to messengers in the superpotential, couplings between $H_u$, $H_d$ and the messengers are induced by standard model D-terms in the scalar potential.  We can associate the operator above, say for the SU(2) adjoint,  to the one loop diagram
\vspace{1cm}
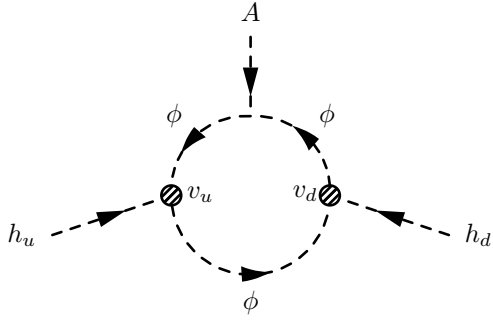
\begin{figure}[H]
	 \centering
	 	 \begin{fmffile}{Aterms1}
	       	 	\begin{fmfgraph*}(150,150)
                   		\fmfleft{i1}
                   		\fmfright{o1}
				\fmftop{t1}
				\fmf{scalar,tension=2}{i1,v1}
				\fmflabel{$h_u$}{i1}
				\fmf{scalar,label=$\phi$,right, tension=1}{v1,v3}
				\fmf{scalar,label=$\phi$,right=0.45,tension=1}{v2,v1}
				\fmf{scalar,label=$\phi$,right=0.45,tension=1}{v3,v2}
                   		\fmf{scalar,tension=2}{t1,v2}
				\fmflabel{$A$}{t1}
                   		\fmf{scalar,tension=2}{o1,v3}
				\fmflabel{$h_d$}{o1}
				\fmfblob{.05w}{v1}
				\fmfblob{.05w}{v3}
				\fmfv{label=$v_u$,label.angle=3.1,label.dist=6}{v1}
				\fmfv{label=$v_d$,label.angle=-3.1,label.dist=-14}{v3}
	        		\end{fmfgraph*}
	    	\end{fmffile}
	\caption{\label{fig:Aterms} Diagrams contributing to A-like terms}
\end{figure}
With a similar contribution resulting from $\overline{\phi}$. There are a equivalent diagram involving S. The diagrams are proportional to an insertion of $\vev{v_u v_d}$, and to the fifth power of standard model gauge coupling.
In models with a messenger parity, the sum of contributions to this parameter is
\be
A_{hhA}= \frac{g_Y^{5} q^3 y^2}{16 \pi^2} \frac{ \vev{v_u v_d}}{m_{\phi}^2}\frac{D}{m_{\phi}} \left(\frac{ gD}{m_{\phi}^2} + \left({\frac{ gD}{m_{\phi}^2}}\right)^3 + . . . \right)
\ee

We may also consider b-terms for Higgses which result from the broken U(1).  The supersoft version of a Higgs b-term was suggested some time ago in the literature, \cite{Fox:2014moa,Fox:2002bu}, and is expressed in operator form as
\be
W= \int d^2 \theta {\delta}_b \frac{W'W'}{\Lambda^2}H_u H_d
\ee
These, again result from messenger loops which involve the D-term coupling of the Higgs's to messengers.  Here the operative loop is

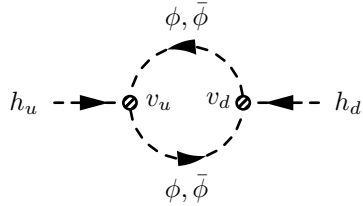
\begin{figure}[H]
	\centering
    		\begin{fmffile}{higgsbterms1}
	        		\begin{fmfgraph*}(100,100)
                  	\fmfleft{i1}
                  	\fmfright{o1}
			\fmf{scalar,tension=1.5}{i1,v1}
			\fmflabel{$h_u$}{i1}
			\fmf{scalar,label=$\phi ,, \bar{\phi}$,right=1,tension=.5}{v1,v2}
			\fmf{scalar,tension=1.5}{o1,v2}
			\fmflabel{$h_d$}{o1}
			\fmf{scalar,label=$\phi ,, \bar{\phi}$,right=1,tension=.5}{v2,v1}
			\fmfblob{.05w}{v1}
			\fmfblob{.05w}{v2}
			\fmfv{label=$v_u$,label.angle=3.1,label.dist=6}{v1}
			\fmfv{label=$v_d$,label.angle=-3.1,label.dist=-14}{v2}
	        		\end{fmfgraph*}
	    	\end{fmffile}
		\caption{\label{fig:bterms} Diagram contributing to b-terms.  Note that this diagram is suppressed by $\vev{v_u v_d}/ \Lambda^2$.}
\end{figure}
\noindent Again we note that either $\phi$ or $\bar{\phi}$ can run in the loop.  This diagram is proportional to the fourth power of the standard model gauge coupling, and $\vev{v_u v_d}$.  Superficially, it appears that the operator is of order $D^2/ \Lambda^2$, thus producing a Higgs sector mass squared the order of ${m_D}^2$, which would be a very large parameter to include in the Higgs potential.  However, we see that this loop has a further suppression proportional to powers of $\vev{v_u v_d}/ \Lambda^2$, like the A-term above.
In the presence of a messenger parity these diagrams sum to
\be
b = \frac{g_Y^{4} q^2}{64 \pi^2} \frac{\vev{v_u v_d}}{m_{\phi}^2} \left( \frac{1}{2}\left(\frac{gD}{m_{\phi}}\right)^2 + \frac{1}{4}\frac{(gD)^4}{m_{\phi}^6} \right) + ...
\ee
We will discuss the suppression of the A and b-terms in the next section.

Finally we note that we find a linear term for S.  Indeed, one may imagine writing a supersoft operator in the superpotential analogous to a Dirac bino term,
\be
W = t_s \frac{W^{'}W^{'}}{\Lambda}S,
\ee
which is of order $D^2/ \Lambda$.   In our messenger model S has a simple trilinear coupling to messengers in the scalar potential. This allows us to draw a tadpole diagram involving S and a messenger loop. In the simplest possible messenger sector we may calculate the leading contribution to this linear term.  Symmetry considerations do not allow this term to arise at order D, therefore the leading term  is of order $D^2$
\be
\frac{\lambda}{16 \pi^2}\frac{m_{\phi}D^2}{m_{\phi}^2},
\ee
as expected.

We note that the messenger completions we have thus far studied are very predictive.  Given certain messenger content and couplings, gaugino masses may be chosen. However, once this occurs, SUSY-breaking contributions to adjoint masses, A-like terms, and the b-term, and linear terms are totally determined.

\subsection{The $\slashed{\mu}$MSSM, new terms, and R-breaking}

So far the models we have been considering do not posses a complete Higgs sectors.  Any such model must generate a sensible $\mu$-term,  demonstrate viable electroweak symmetry breaking, and predict a Higgs mass in line with current measurements.  We now will consider an extension of R-symmetric models which generates a $\mu$-term, and holds the possibility of predicting a 125 GeV Higgs mass, through increased tree level quartic couplings, and new loop effects.

This class of models has large superpotential coupling between the Higgs fields and the adjoint scalars.  These couplings, taken together with messenger couplings give rise to new quartic couplings, as well as to several interesting one loop effects, Higgs soft masses,  b-terms, and A-like terms. All of these new terms will drastically effect the Higgs-potential.

Consider the following superpotential
\be
W= \sqrt{2}\lambda_A H_u A H_d + \lambda_SS H_u H_d  +\sqrt{2} y_A  \phi A\overline{\phi} + y_S  S \phi \overline{\phi} + m_{\phi} \phi \overline{\phi},
\ee
where A is an $SU(2)$ adjoint, S a standard model singlet, and $\phi$ and $\overline{\phi}$ are the messengers which couple to the adjoint and singlet
and have a supersymmetric mass.  We first note that the above potential generates several separate contributions to Higgsino masses.  This was christened the $\slashed{\mu}$MSSM by Nelson et al and was a proposal for the solution of the $\mu$ problem.  By inserting just $H_u$ and $H_d$ vevs we find separate Dirac masses for the up and down type Higgsino.
\be
\sqrt{2}\lambda_A \vev{v_u} \Psi_A \Psi_{hd}+ \lambda_S \vev{v_u} \Psi_S \Psi_{hd} + \sqrt{2}\lambda_A \vev{v_d} \Psi_{hu } \Psi_A + \lambda_S \vev{v_d} \Psi_{hu} \Psi_S
\ee
Thus producing two R-symmetric $\mu$-terms for Higgsinos. The SU(2) adjoint vev must remain very small to be in-line with precision constrains, however if the S field acquires a vev, there is yet another contribution to the $\mu$-term.  The result for fermions, is an electroweakino mass matrix with non-trivial mixing between Higgsinos and adjoint/singletinos.

Lets now explore the consequences of  R-breaking in the superpotential.  Preserving R-symmetry in the models we are studying has required the chiral adjoints to have R-charge 0.  We see however,  R-symmetry is broken during electroweak symmetry breaking. As an example, we may look at the effective superpotential for fields S, $H_u$, $H_d$.

\be
W &\supset& \lambda_S S H_u H_d + {\delta}_S \frac{W'W'}{\Lambda^3}S H_u H_d + {\delta}_b \frac{W'W'}{\Lambda^2} H_u H_d
\ee

Here the Higgs $\mu$ term comes from trilinear terms with adjoints.  With the R-charge assignment of S set to zero, we see that Higgses must carry R-charge. (This is also case if we force the $\mu$-term to arise from a fundamental supersymmetric mass term $\mu H_u H_d$.) The charge of $H_u H_d$, in fact, must be 2. We see that the induced b and A-terms ($\delta_b$ and $\delta_S$ respectively) do not allow R-symmetry to be preserved and are R-breaking, each of R charge 4.

When electroweak symmetry breaking occurs, the Higgses will get R-breaking vevs. We have seen that the R-breaking diagrams in the Higgs sector, A-like terms and Higgs b-terms, are proportional to the R-breaking parameter $\vev{v_u v_d}/\Lambda^2$. We perhaps may think of the R-breaking operators as each coming with an R-breaking spurion to soak up the excess R-charge. This increases the dimension of the operator which we would then expect these operators to be further suppressed by powers of $\Lambda$.

The Higss-adjoint superpotential terms generate new contributions to the scalar potential for cross terms involving A:
\begin{eqnarray}
V &\supset& y^2|A|^2(|\phi|^2+|\bar\phi|^2)+ym_{\phi}(|\phi|^2+|\bar\phi|^2)(A+A^*)+m_{\phi}^2(|\phi|^2+|\bar\phi|^2) \nonumber \\
&+&y^2|\phi\bar{\phi}|^2+y\lambda_A(\phi\bar\phi H_u^* H_d^* + \text{h.c.})
\end{eqnarray}
with similar terms involving S.
One immediate consequence of this scalar potential are extra couplings between Higgs and the messengers scalars.  We see that there is now a $\mu$ term for messengers proportional to higgs vevs.  This will enhance contributions in this model to the A-like and b-terms mentioned above.

We may now write a new one-loop contribution to the A-like operators, still proportional to $\vev{v_u v_d}$. Below we show a contribution to the A-term of the SU(2) adjoint.  There are an equivalent set of diagrams for the singlet.

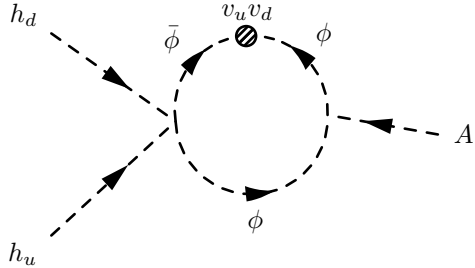
\begin{figure}[H]
	 \centering
	 	 \begin{fmffile}{Aterms}
	       	 	\begin{fmfgraph*}(150,150)
                   		\fmfleft{i1,i2,i3,i4,i5}
                   		\fmfright{o1,o2,o3,o4,o5}
				\fmftop{t2}
				\fmf{scalar,label=$\phi$,right, tension=0.75}{v1,v3}
				\fmf{scalar,label=$\bar{\phi}$,left=0.45,tension=.85}{v1,v2}
				\fmf{scalar,label=$\phi$,right=0.45,tension=.85}{v3,v2}
				\fmf{phantom,tension=.9}{v1,v3,o3}
                   		\fmf{scalar,tension=1.5}{i2,v1}
				\fmflabel{$h_u$}{i2}
                   		\fmf{scalar}{i4,v1}
				\fmflabel{$h_d$}{i4}
				\fmf{scalar, tension=2}{o3,v3}
				\fmflabel{$A$}{o3}
				\fmf{phantom,tension=1.4}{t2,v2}
				\fmfblob{.05w}{v2}
				\fmflabel{$v_uv_d$}{v2}
	        		\end{fmfgraph*}
	    	\end{fmffile}
\end{figure}	
\noindent With equivalent diagram $\phi \rightarrow \overline{\phi}$.  Now the diagrams are proportional to the fourth powers of large Yukawa-like couplings $y$ instead of gauge couplings.  Again, the order of SUSY breaking is $D^2/ \Lambda^3$. With
\be
A_{hhA}= \frac{y^3\lambda }{8 \pi^2} \frac{  \vev{v_u v_d} }{m_{\phi}} \left(\frac{ gD^2}{m_{\phi}^2} + {\frac{ gD}{m_{\phi}^4}}^4 + . . . \right).
\ee
There is an additional diagram equivalent to \fig{fig:Aterms} with the $A|\phi|^2$ vertex now coming from the $F-term$ potential instead of the $D-term$ potential.  This contribution goes as
\be
A_{hhA}= \frac{y^2v_s+ym_{\phi}}{8 \pi^2}\frac{ \vev{v_u v_d}}{m_{\phi}^2}\left( \left({\frac{ gD}{m_{\phi}^2}}\right)^2 + . . . \right)
\ee

The b-term also gets a new contribution from the same Higgs-messenger interaction:
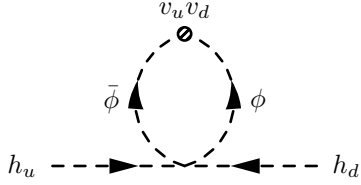
\vspace{1cm}
\begin{figure}[H]
	\centering
    		\begin{fmffile}{higgsbterms}
	        		\begin{fmfgraph*}(100,100)
                  	\fmfleft{i1}
                  	\fmfright{o1}
			\fmftop{t1}
                   	\fmf{scalar}{i1,v1}
			\fmflabel{$h_u$}{i1}
                   	\fmf{scalar,label=$\phi$,right=0.75,tension=0}{v1,v2}
			\fmf{scalar,label=$\bar{\phi}$,left=0.75,tension=0}{v1,v2}
			\fmf{phantom,tension=1}{t1,v2}
                   	\fmf{scalar}{o1,v1}
			\fmflabel{$h_d$}{o1}
			\fmfblob{.05w}{v2}
			\fmflabel{$v_uv_d$}{v2}
	        		\end{fmfgraph*}
	    	\end{fmffile}
\end{figure}
\noindent which is also proportional to the fourth power of the Yukawa-like coupling y, and the R breaking combination $\vev{v_u v_d}$.

\be
b = \frac{\lambda^2 y^2 }{32 \pi^2}\frac{\vev{v_u v_d}}{m_{\phi}^2}\left(\frac{1}{3} \frac{g^2D^2}{m_{\phi}^4} + \frac{1}{10}\frac{g^4D^4}{m_{\phi}^8}\right)+ ...
\ee

We note that we may also write general tri-linear A-like terms for Adjoint fields which are R-preserving thus unsuppressed by Higgs vevs.  In the electroweak sector, these operators may be written in the super-potential

\be
W_{\text{trilinear}} = \zeta_S\frac{W^{'}W{'}}{\Lambda^3}S^3 +\zeta_{AS}\frac{W^{'}W{'}}{\Lambda^3}\text{Tr}[AA]S + \zeta_A\frac{W^{'}W{'}}{\Lambda^3}d^{abc}[A^aA^bA^c]
\ee
These are quite analogous to the Higgs sector A-like terms and can be computed in a similar diagrammatic fashion.



\section{The Higgs Sector }
\subsection{Tree Level}
We will now consider electroweak symmetry breaking in a model with general operators.  This Higgs sector is quite complex, and variations of the Higgs potential in R-symmetric models have been recently been studied \cite{Bertuzzo:2014bwa,Benakli:2012cy}. We will assumed that there is no explicit R symmetry breaking,  all terms which violate R symmetry will be generated spontaneously and we expect these should be suppressed with respect to R-symmetric terms.  Our low energy superpotential will contain the Dirac-gaugino masses for the elctroweak sector as well electroweak adjoints coupling directly to the Higgs \cite{Nelson:2002ca}.
 The low energy superpotential will then take the form
\begin{equation}
W= y_A\frac{W^{'}WA}{\Lambda^2}+ y_S\frac{W^{'}WS}{\Lambda^2}\sqrt{2}\lambda_A H_uAH_d+\lambda_S SH_uH_d+ \text{Yukawa terms}.
\end{equation}
Where A is the SU(2) adjoint and S is a singlet. Here the SU(2) adjoint A has been parameterized so that
\begin{eqnarray}
A= \frac{\sigma^a}{2}A^a =\frac{1}{\sqrt{2}}\left( \begin{array}{cc}
A^0 & \sqrt{2}A^+ \\
\sqrt{2}A^- & -A^0 \\
\end{array}\right)
\end{eqnarray}


All A-terms, b-terms and SUSY breaking scalar masses-both holomorphic and non-holomorphic mass terms for the Higgs fields and the adjoints- will be included in the scalar potential.

\begin{eqnarray}
V_{\text{soft}} &=& m_u^2|H_u|^2+m_d^2|H_d|^2+m_{S}^2|S|^2+2m_{A}\text{Tr}|A|^2 \nonumber \\ 
&+&\left(bH_uH_d + b_A\text{Tr}A^2+\frac{1}{2}b_S S^2+\text{h.c.}\right) \nonumber \\ 
&+&(A_{AH} A  H_u H_d + A_{SH} S H_u H_d + A_S S^3  + A_{AS}\text{Tr}[AA]S + A_A AAA + h.c.) \nonumber \\
&+& (t_S S + h.c.) .
\end{eqnarray}
Contributions to these parameters come from the SUSY breaking D-terms, as discussed above. We have written the scalar potential in general terms to account for a general structure within the messenger sector. One may invoke sectors with multiple set of messengers and a breaking of messenger parity, see for example \cite{Csaki:2013fla}.  Or one may invoke messenger sectors as in \cite{Carpenter:2010as} that  produce additional log divergent, R-preserving scalar masses which follow from two-loop gauge mediated contributions as in \cite{Poppitz:1996xw}.

\subsection{Calculating and Minimizing the scalar Potential}
Contribution to the Higgs scalar potential comes from three pieces, F-terms, D-terms, and
the soft SUSY breaking terms.  Recall that the presence of the supersoft operators shift the
standard model D-terms
\begin{eqnarray}
D_2\rightarrow m_D(A+A^{\dagger})+\Sigma_i gQ_i^\dagger TQ_i \\
D_1 \rightarrow m_D(S+S^{\dagger})+\Sigma_i gq_iQ_i^\dagger Q_i \nonumber
\end{eqnarray}

We assume the minimum does not break U(1)$_{EM}$, thus we take the vevs of the charged fields to be zero $h_u^+=h_d^-=A^+=A^-=0$.
The full neutral scalar potential is then
\begin{eqnarray}
V&=& \frac{g^2+g'^2}{8}(|h_u^0|^2-|h_d^0|^2)^2+(\lambda_A^2+\lambda_S^2)|h_u^0|^2|h_d^0|^2+[\lambda_A ^2|A_0|^2+\lambda_S^2|S|^2+\lambda_A\lambda_S(A_0S^*+\text{h.c})](|h_u^0|^2+|h_d^0|^2) \nonumber \\
&-&\frac{1}{\sqrt{2}}gm_{D_A}(A_0+A_0^*)(|h_u^0|^2-|h_d^0|^2)+\frac{1}{2}g'm_{D_S}(S+S^*)(|h_u^0|^2-|h_d^0|^2) \nonumber \\
&+& 2(m_{A}^2+m_{D_A}^2)|A_0|^2+(m_{S}^2+m_{D_S}^2)|S|^2+m_{h_u}^2|h_u^0|^2+m_{h_d}^2|h_d^0|^2\nonumber \\
&+&(b_A +m_{D_A}^2)(A_0^2+A_0^{*2})+\frac{1}{2}(b_S+m_{D_S}^2)(S^2+S^{*2})-(bh_u^0 h_d^0 + h.c.) \nonumber \\
&+& (A_{AH} A_0  h_u h_d + A_{SH} S h_u h_d + A_S S^3  + A_{A S} S |A_0|^2 + A_A A_0^3 + h.c.)+ (t_S S + h.c.)
\end{eqnarray}

This potential is similar to the general potential studied by \cite{Benakli:2012cy}. As we
do not want to introduce any additional CP violation we take all fields and parameters to be real.
The consistency of these assumptions were checked numerically.  In general we see the linear term for S leads to
 $V\left(\langle h_u^0\rangle,
\langle h_d^0\rangle,\langle A^0\rangle,\langle S\rangle \ne 0\right)<V\left(\langle h_u^0\rangle=
\langle h_d^0\rangle=\langle A^0\rangle=\langle S\rangle=0\right)$ .

The tree level behavior of this potential is quite different from the MSSM.  While the Higgs quartic of the MSSM is generated only by the SM D-terms, here, the addition of the supersymmetric trilinear couplings gives additional tree level contributions to the quartic.  We thus expect that a choice of large couplings for $\lambda_A$ and/or $\lambda_S$ will lead to an enhancement in the tree-level Higgs mass.  In the next section we will calculate new one-loop contributions to the Higgs mass due to the new adjoint couplings.

We now consider the spectra of the Higgs sector given certain parameter inputs.  We consider a points with TeV-scale Dirac gaugino mass as would be typical of Supersoft models.  We note that in gauge-mediated completions of Supersoft as we have considered here, we expect both Dirac gaugino masses and holomorphic/non-holomorphic adjoint scalar masses to be large and of the same order. 
\vspace{.5cm}
\begin{table}[H]
\centering
\begin{tabular}{| c | c |}
\hline
tan$\beta$ & 10 \\
$\lambda_A$ &  $0.021$ \\
$\lambda_S$ & $-1.3$   \\
$v_S$            & 180 \\
$m_{D_A}$    & $2$ TeV\\
$m_{D_S}$    & $ 1.5$ TeV\\
$m_A^2$        & $2.5*10^7$ \\
$b$                 & $4.2*10^4$\\
$b_S$             & $-5*10^5$ GeV$^2$ \\
$b_A$            & 0 \\
$t_S$              & $-1.5*10^9$ GeV$^3$\\
$A$-like terms              & 0\\
\hline
\end{tabular}
\end{table}
\vspace{.5cm}
At tree level the neutral higgs spectrum is ($89, 648, 3442, 8124$) GeV. The lightest Higgs is almost entirely $h_u$.  The pseudo-scalar spectrum is $650, 2868, 7071$ GeV. With charged Higgses $612, 7071, 8124$ GeV.  The S soft mass at this point is 2770 GeV.  We note that the high-mass point has parametric scaling of parameters roughly in line with the calculations from a gauge mediated completion.  Here both the Dirac masses and adjoint scalar masses are of the same order, TeV in scale, with a linear S term also of order $(TeV)^3$.  We expect parametrically small A-like terms for Higgses, of order $D^2 v^2/\Lambda^5$ so we may neglect them in this case.  And a Higgs b-term suppressed by powers of R-breaking.  The vev of the SU(2) adjoint remains very small, as required to be in-line with electroweak precision constraints.

\subsection{Loop corrections to Higgs mass from $W=\lambda_A H_dAH_u+\lambda_S SH_dH_u$}
Once we introduce trilinear superpotential couplings of the Higgs to the $SU(2)$ adjoint and the singlet, we find large loop corrections to the Higgs soft mass.   The one-loop diagrams contributing to the Higgs mass are very similar to those one obtains from top/stop loops in the MSSM, and are shown below.
\vspace{.25cm}
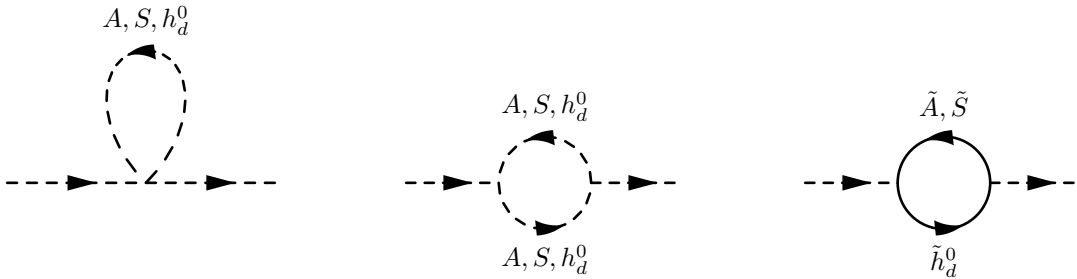
\begin{figure}[H]

         \begin{subfigure}[b]{.3\textwidth}
         \centering
	    \begin{fmffile}{4point}
	        \begin{fmfgraph*}(105,105)
                   \fmfleft{i1}
                   \fmfright{o1}
                   \fmf{scalar}{i1,v1}
                   \fmf{scalar,label=$A,,S,,h_d^0$,right,tension=.7}{v1,v1}
                   \fmf{scalar}{v1,o1}
	        \end{fmfgraph*}
	    \end{fmffile}
           \end{subfigure}
~
         \begin{subfigure}[b]{.3\textwidth}
         \centering
	    \begin{fmffile}{3point}
	        \begin{fmfgraph*}(105,105)
                   \fmfleft{i1}
                   \fmfright{o1}
                   \fmf{scalar}{i1,v1}
                   \fmf{scalar,label=$A,,S,,h_d^0$,right,tension=.5}{v2,v1,v2}
                   \fmf{scalar}{v2,o1}
	        \end{fmfgraph*}
	    \end{fmffile}
           \end{subfigure}
~
        \begin{subfigure}[b]{.3\textwidth}
        \centering
	    \begin{fmffile}{3pointFermion}
	        \begin{fmfgraph*}(105,105)
                   \fmfleft{i1}
                   \fmfright{o1}
                   \fmf{scalar}{i1,v1}
                   \fmf{fermion,label=$\tilde{A},,\tilde{S}$,right,tension=.5}{v2,v1}
                   \fmf{fermion,label=$\tilde{h}_d^0$,right,tension=.5}{v1,v2}
                   \fmf{scalar}{v2,o1}
	        \end{fmfgraph*}
	    \end{fmffile}
	  \end{subfigure}
	  \caption{\label{fig:HiggsMassCorr} Corrections to Higgs mass due to new quartic interactions.}
\end{figure}
\noindent To calculate corrections to the Higgs mass we invoke the one-loop effective potential,
\be
V = V_0 + V_{CW}
\ee
Where $ V_{CW}$ is the Coleman-Weinberg potential
\be
V_{CW}= \frac{1}{64 \pi^2} Str[\mathcal{M}^4]\left( \log \frac{\mathcal{M}^2}{Q^2} \right)-\frac{3}{2})
\ee
and $V_0$ is the tree level potential, Q is the renormalization scale, and the supertrace is taken over the fields which couple to the Higgs.  We calculate induced corrections to the Higgs quartic coupling when heavy adjoints are integrated out.   We present results in the limit that $m_D$ is large, $>>$ $v_h$, with adjoint soft masses the same order as the Dirac mass. This region of parameter space is appropriate for gauge mediated supersoft models, as we have demonstrated.  Typical Higgs mass corrections may be quite large.  Couplings between the Higgsses and the adjoints may be large, $\lambda_S$ is typically the size of the top Yukawa, while the adjoint scalars are heavy.

Mass corrections are thus equal in size, larger than stop mass corrections in the MSSM.  For our tree-level point above, we find corrections to the Higgs quartic $\delta \lambda_h \sim .055$ leading to a Higgs mass correction $\delta_{m_u^2}\sim (54\text{GeV})^2$.  Including these one loop corrections, the $89$GeV tree-level Higgs mass is pushed up to $\sim 143$GeV.  Thus we note an interesting feature of models with typical  Supersoft-parameters, which is that they may, in fact,  over-predict the Higgs mass in many regions of parameter space.  We note that the bottom-type Higgs quartic is also shifted significantly, however this does not greatly effect the lightest Higgs mass in our typical parameter space as the bottom Higgs vev remails small.

\section{Conclusions}
We have presented a formalism for calculating general one-loop SUSY breaking parameters that arise in models with Dirac gauginos.  We have studied models which are completed with a messenger sector, where messengers are  charged under the standard model gauge groups as well a hidden sector U(1).  We have used an expansion technique to calculate Dirac gaugino masses, as well as holomorphic and non-holomorphic SUSY breaking masses of chiral standard model adjoint scalars to arbitrary order in the SUSY breaking mass parameter D.  This technique may be applied to sectors with any number of messengers, these sectors may or may not have a messenger parity.

We have also computed one loop SUSY breaking contributions to adjoint scalar masses, gaugino masses, Higgs b-terms, A-like parameters involving the adjoint scalars and Higgs fields, and singlet linear terms. We have shown the relation between these parameters and SUSY-breaking adjoint and gaugino masses for given messenger sectors.  We have analyzed these terms with and without extra superpotential coupling between Higgses and chiral adjoints, the `$\slashed{\mu}$MSSM' type models. In general we have found holomorphic and non-holomorphic adjoint masses to be of order $m_D$ in models without messenger parity.  We have also found suppression of R-breaking b-terms and A-like terms and generally large contributions to the linear term S.

We have explored the phenomenological prospects for  models large Dirac masses with regard to Higgs sector parameters.  We find fairly acceptable electroweak symmetry breaking minima for supersoft-like models with large Dirac gaugino masses.  The tree level Higgs masses vary tremendously over the parameter space, with many points actually over-predicting the tree level Higgs mass.  We then calculated large one-loop corrections to Higgs masses caused by Higgs couplings to chiral adjoints.

Predictions in these models are very dependent on the messenger sectors.  We see that several SUSY-breaking parameters including holomorphic adjoint scalar masses are greatly changed by the absence or presence of messenger parity as models with messenger parity raise the order at which operators appear.  Once the messenger sector is set, there are definite relations between SUSY breaking parameters.  Many relations between the SUSY-breaking parameters among R-preserving operators are fixed simply by setting supersymmetric messenger masses and their coupling to adjoints.  Dialing the Higgs-adjoint couplings gives only a small amount of freedom, mostly to R-breaking parameters whose magnitude is always in any case.  Given this predictivity, it is a good avenue for further study to determine if specific UV models can remain phenomenologically viable in all sectors.  This is especially true in the Higgs sector were we have found linear terms for singlets, large b-terms for adjoints, and small b-terms for Higgses. Our modestly successful test point did not follow from a complete theory.  Achieving further independence of the SUSY breaking parameters, will require models with more structure.

One possibility for further model building is to invoke Dirac gaugino models with multiple SUSY breaking sectors.  Such models may have quite rich interesting phenomenology, see for example \cite{Carpenter:2005tz}.   One may, for example, have a SUSY breaking sector with broken U(1) gauge symmetry, in addition to a sector with F-term SUSY breaking allowing for F-term SUSY breaking through new messengers with Yukawa mediation instead of gauge interactions. This could induce new one loop A-like terms and adjoint soft masses. R-breaking, however would be sequestered from gauginos at leading order.  These models will be topics of further study.  Another interesting avenue would be 'retrofitted' models,  building mass parameters in the superpotential that are due to gaugino condensation\cite{Dine:2006gm}. (See \cite{Carpenter:2008rj} for an example of a model combining these terms with  supersoft operators).  This set-up  would allow, for example, R-symmetric $\mu$ terms which assign the Higgs W charge 0. This may ease the suppression of A-like terms and b-terms in the Higgs potential.

{\bf Acknowledgments}
The authors would like to thank Yuri Shirman and Stuart Raby for helpful discussions.  JG would also like to thank Florian Staub for helpful discussions.  We thank the physics department at Ohio State University for funding this work.

\section*{Appendix}
In general messenger sectors which may not have a messenger parity, we give the leading contributions of each messengers to various parameters.  Recall that a general messenger sector is given by
\be
W_\phi= m_{ij} \Phi_i\overline{\Phi_j}+ y_{ij}\overline{\Phi_i}A\Phi_j+ y_{ij}^{'}\overline{\Phi_i}S\Phi_j
\ee
Where the masses $m_{ij}$ may be generated most simply as the vevs of a set of chiral fields.
\be
m_{ij}= \lambda_{ijk}Y_k
\ee
Thus most generally, $\phi_i\overline{\phi_j}$ may have different charges under the hidden sector gauge group depending on the various charged of the fields $Y_k$.  As the chiral adjoints are a singlets under the hidden sector gauge group, the pair of messengers that couple to them $\phi_i$ and $\overline{\phi_j}$ must have opposite U(1) charges, or else the coupling  $y_{ij}$ must be $0$.

1. \textbf{$t_s$ contributions} \\

We now write a general expression for terms contributing to the SUSY-breaking linear term of the singlet S.  In order to generate this term, the fields $\phi_i$ and $\overline{\phi_j}$ must have equal and opposite U(1) charge.  In addition they must possess a non-zero mass term $m_{ij}$ which mixes these two fields.
Fermion loops then involve the pair of fermions $\phi_i\overline{\phi_j}$ their  Dirac mass insertion
\be
\frac{1}{4 \pi^2}\left(\Lambda^2-\left(m_{ij}\right)\text{ln} \left(\frac{\Lambda^2 +m_{ij}^2}{m_{ij}^2}\right)\right)
\ee
Each messenger $\phi_i$ or $\overline{\phi_j}$ gives one loop scalar contribution, where $\phi_i$ and $\overline{\phi_j}$ have opposite sign D-term masses.
\be
 \frac{1}{8 \pi^2}\left(\Lambda^2-(m_{\phi_i}^2-gD)\text{ln} \left(\frac{\Lambda^2 +m_{\phi_i}^2 - gD}{{m_{\phi_i}}^2 -gD}\right)\right)
\ee

The quadratic divergence cancels pair by pair for messengers that couple to the singlet.  \\

2. \textbf{b terms} \\

For b-terms proportional to SM D terms we will give the leading terms. Each loop contributing to this process contains a single messenger running in the loop.  Each messenger loop containing a $\phi_i$  gives a contribution
\be
b = \frac{g^{'4} q^2 }{128 \pi^2}\frac{\vev{v_u v_d}}{m_{\phi_i}^2}\left(gD + \frac{g^2D^2}{2 m_{\phi_i}^2} + . . .  \right)
\ee
Where $g^{'}$ is the SM U(1) gauge coupling and q is the messenger charge under the hidden-sector $U(1)^{'}$. One must sum the contribution of all messengers(fundamentals and anti-fundamentals) that couple to the Higgses through a Standard Model D-term.
\\
For b-terms that arise in $\slashed{\mu}$MSSM models, messenger loops contain pairs of messengers $\phi_i\overline{\phi_j}$ which couple to the chiral adjoints with coupling $y_{ij}$.  The messenger pairs in the loops must have equal and opposite $U(1)^{'}$ charges.


3. \textbf{A terms} \\

 We now discuss the generation of A-like terms arising from SM D terms. In order to generate this term, messengers pairs which couple to the chiral adjoints $\phi_i$ or $\overline{\phi_j}$ must have a non-zero mass term $m_{ij}$.  Here $\phi_i$ and $\overline{\phi_j}$must have equal and opposite $U(1)^{' charges}$. There are two loops to sum for each messenger pair that couples to the adjoint. Each loop contains a single messenger $\phi_i$ or $\overline{\phi_j}$.  Each loop may be expressed as
\be
\frac{g'^5 q^3}{16 \pi^2}\frac{\vev{v_u v_d}}{m_{\phi_i}^2}m_D\left(\frac{gD}{m_{\phi_i}^2} + \frac{g^2D^2}{m_{\phi_i}^4}... \right)
\ee

Where $g^{'}$ is the SM U(1) gauge coupling and q is the messenger charge under the hidden-sector $U(1)^{'}$. We must sum the messengers in pairs, adding up all pairs that couple to the chiral adjoints .
\\

%



\begin{thebibliography}{9}

\bibitem{Chatrchyan:2012ufa}
  S.~Chatrchyan {\it et al.}  [CMS Collaboration],
  Phys.\ Lett.\ B {\bf 716}, 30 (2012)
  [arXiv:1207.7235 [hep-ex]].

\bibitem{Aad:2012tfa}
  G.~Aad {\it et al.}  [ATLAS Collaboration],
  Phys.\ Lett.\ B {\bf 716}, 1 (2012)
  [arXiv:1207.7214 [hep-ex]].


\bibitem{Fayet:1978qc}
  P.~Fayet,
  Phys.\ Lett.\ B {\bf 78}, 417 (1978).



\bibitem{Hall:1990hq}
  L.~J.~Hall and L.~Randall,
  Nucl.\ Phys.\ B {\bf 352}, 289 (1991).

\bibitem{Kribs:2012gx}
  G.~D.~Kribs and A.~Martin,
  Phys.\ Rev.\ D {\bf 85}, 115014 (2012)
  [arXiv:1203.4821 [hep-ph]].

\bibitem{Fox:2014moa}
  P.~J.~Fox, G.~D.~Kribs and A.~Martin,
  Phys.\ Rev.\ D {\bf 90}, no. 7, 075006 (2014)
  [arXiv:1405.3692 [hep-ph]].

\bibitem{Fox:2002bu}
  P.~J.~Fox, A.~E.~Nelson and N.~Weiner,
  ``Dirac gaugino masses and supersoft supersymmetry breaking,''
  JHEP {\bf 0208}, 035 (2002)
  [hep-ph/0206096].

\bibitem{Dine:1993yw}
  M.~Dine and A.~E.~Nelson,
  Phys.\ Rev.\  D {\bf 48}, 1277 (1993)
  [arXiv:hep-ph/9303230].

\bibitem{Dine:1994vc}
  M.~Dine, A.~E.~Nelson and Y.~Shirman,
  Phys.\ Rev.\  D {\bf 51}, 1362 (1995)
  [arXiv:hep-ph/9408384].

\bibitem{Nelson:2002ca}
  A.~E.~Nelson, N.~Rius, V.~Sanz and M.~Unsal,
  ``The Minimal supersymmetric model without a mu term,''
  JHEP {\bf 0208}, 039 (2002)
  [hep-ph/0206102].

\bibitem{Meade:2008wd}
  P.~Meade, N.~Seiberg and D.~Shih,
  Prog.\ Theor.\ Phys.\ Suppl.\  {\bf 177}, 143 (2009)
  [arXiv:0801.3278 [hep-ph]].

\bibitem{Benakli:2008pg}
  K.~Benakli and M.~D.~Goodsell,
  Nucl.\ Phys.\  B {\bf 816}, 185 (2009)
  [arXiv:0811.4409 [hep-ph]].

\bibitem{Seiberg:2008qj}
  N.~Seiberg, T.~Volansky and B.~Wecht,
  JHEP {\bf 0811}, 004 (2008)
  [arXiv:0809.4437 [hep-ph]].


\bibitem{Dine:2007dz}
  M.~Dine and J.~D.~Mason,
  Phys.\ Rev.\ D {\bf 78}, 055013 (2008)
  [arXiv:0712.1355 [hep-ph]].

\bibitem{Carpenter:2010as}
  L.~M.~Carpenter,
  JHEP {\bf 1209}, 102 (2012)
  [arXiv:1007.0017 [hep-th]].

\bibitem{Csaki:2013fla}
  C.~Csaki, J.~Goodman, R.~Pavesi and Y.~Shirman,
  ``The $m_D-b_M$ Problem of Dirac Gauginos and its Solutions,''
  Phys.\ Rev.\ D {\bf 89}, 055005 (2014)
  [arXiv:1310.4504 [hep-ph]].

\bibitem{Giudice:1997ni}
  G.~F.~Giudice and R.~Rattazzi,
  Nucl.\ Phys.\ B {\bf 511}, 25 (1998)
  [hep-ph/9706540].

\bibitem{Nelson:2015cea}
  A.~E.~Nelson and T.~S.~Roy,
  arXiv:1501.03251 [hep-ph].

\bibitem{Benakli:2012cy}
  K.~Benakli, M.~D.~Goodsell and F.~Staub,
  JHEP {\bf 1306}, 073 (2013)
  [arXiv:1211.0552 [hep-ph]].

\bibitem{Poppitz:1996xw}
  E.~Poppitz and S.~P.~Trivedi,
  Phys.\ Lett.\ B {\bf 401}, 38 (1997)
  [hep-ph/9703246].

\bibitem{Bertuzzo:2014bwa}
  E.~Bertuzzo, C.~Frugiuele, T.~Gregoire and E.~Ponton,
  arXiv:1402.5432 [hep-ph].



\bibitem{Carpenter:2005tz}
  L.~M.~Carpenter, P.~J.~Fox and D.~E.~Kaplan,
  ``The NMSSM, anomaly mediation and a Dirac bino,''
  hep-ph/0503093.

\bibitem{Dine:2006gm}
  M.~Dine, J.~L.~Feng and E.~Silverstein,
  Phys.\ Rev.\ D {\bf 74}, 095012 (2006)
  [hep-th/0608159].

\bibitem{Carpenter:2008rj}
  L.~M.~Carpenter,
  JHEP {\bf 1105}, 069 (2011)
  [arXiv:0809.0026 [hep-ph]].



















\end{thebibliography}
\end{document}